# Edible Resonators

Ganapathy Saravanavel[1], Sanjay John[1], Gunashekar K. R.[1], Gwenhivir Wyatt-Moon[2], Andrew Flewitt[2] and Sanjiv Sambandan[1,2,3]

[1] Department of Instrumentation and Applied Physics, Indian Institute of Science, India 560012
[2] Department of Engineering, University of Cambridge, Cambridge, United Kingdom CB30FF
[3] Department of Electronic Systems Engineering, Indian Institute of Science, India 560012

E-mail: sanjiv@iisc.ac.in, ss698@cam.ac.uk

**Abstract**

The development of green, biodegradable electronics that degrade soon after their function is accomplished is a significant contribution to healthcare where edible electronics can be used for diagnosis. An important component of such systems is a resonator. Here we present the design and development of inductor-capacitor (LC) circuits based on a form of sugar called isomalt. Using its low roughness and mechanical properties, we demonstrate resonators that use capacitors, planar inductors, helical inductors and inductors with an edible core, all fabricated from isomalt.

## 1. Introduction

The recent years have seen an increase in the development of electronics on flexible and non-planar substrates [1]-[6]. With new technologies and low temperature fabrication, this has led to possibilities of large scale manufacturing. This however has come at the cost of reliability and there have been several technologies developed to counter this problem [7]-[9].

One sector of application of flexible electronics has been in the space of healthcare [10]-[14]. In this space, the development of conformable, soft, bio-compatible, and bio-degradable electronics offer interesting possibilities [15]-[17]. More recently, an interesting thread in healthcare has been the possibility of electronics that are edible and digestible [18]-[34]. These electronic devices are built materials that are non-toxic (or are of low trace amounts well below toxicity limits). Such devices promise decentralized diagnostics.

With these possibilities in mind, in this work we discuss the development of electronics on isomalt substrates, which is a form of sugar. More specifically, we discuss the design of inductors, capacitors and passive resonators with the goal of developing oscillators and communication circuits. The toxicity limit of all materials used is well below thresholds for a child.

## 2. Results

### 2.1 *Substrate materials*

From the materials perspective, there exist several environmentally friendly options for use as a substrate for electronic devices, e.g. cellulose, collagen, chitin etc. In this work we explore the use of two forms of sugar called isomalt and sugarin both of which are commonly used for baking. Isomalt is a sugar substitute that is commonly available as an odorless, white, crystalline powder with reasonably low hygroscopicity. At high temperatures, isomalt becomes viscoelastic and can be molded. Although widely used in sugar art and as an artificial sweetener, its scientific application was explored by the development of monolithic multilayer micro fluidics via sacrificial molding of 3D printed isomalt [35], [36].

*2.1.1. Fabrication and Metallization.* The process flow for the preparation of isomalt substrates is shown in **Fig. 1a**. Isomalt pellets were heated to 200°C. After cooling to 120 °C, the viscous melt was sandwiched between smooth surfaces e.g. made of polydimethyl siloxane (PDMS), and compressed to the desired thickness. The surface roughness of the prepared isomalt substrates was found by atomic force microscopy to be about 4.0 nm to 20.0 nm (**Fig 1b**). Fourier transform infrared spectroscopy of the substrate revealed the presence of C-O, C-C, C=O, C-H, and O-H, bonds respectively (**Fig. 1c**). Isomalt substrates were susceptible to moisture and dissolved rapidly in water (**Fig. 1d**). While this is a useful feature with regards to biodegradable electronics, it is undesirable with regards to electronic device fabrication as exposure to moisture would increase surface roughness. To make isomalt substrates less susceptible to moisture, the fabricated isomalt substrate was dip-coated with the polymer edible glaze (Sugarin) as described in **Fig. 1e**.

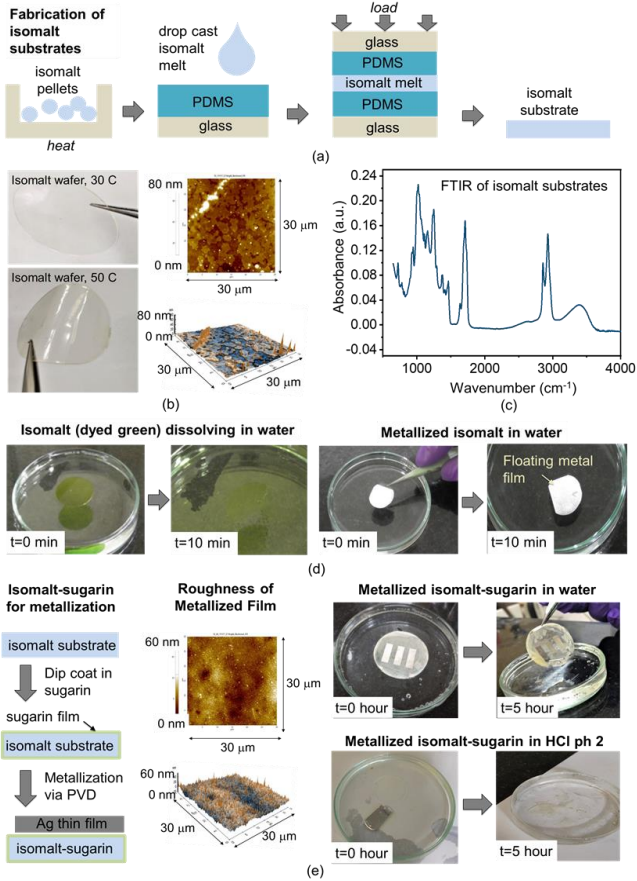

**Figure 1.** Fabrication of isomalt substrates (a) Illustration of the fabrication process flow of isomalt substrates (b) Surface morphology of isomalt substrates (c) FTIR spectroscopy of isomalt substrates (d) Solubility of isomalt and metallized isomalt in water (e) Illustration of the process of metallization of isomalt-sugarin substrates, surface morphology of metallized isomalt-sugarin substrates and solubility of metallized isomalt-sugarin substrates in water and HCl (pH 2).

Both isomalt and isomalt-sugarin substrates were vacuum compatible with low outgassing rate up to $10^{-5}$ mbar and suitable for metallization via dc magnetron sputtering. Patterning was achieved by the use of stainless steel shadow masks that achieved a spatial resolution of 100 μm. The surface roughness after 150 nm thick silver metallization on isomalt-sugarin substrates via dc magnetron sputtering was found to be 10 nm to 20 nm (**Fig. 1e**).

*2.1.2. Mechanical Properties of Isomalt.* Flexural strength measurements of isomalt substrates (2 cm diameter and 1mm thickness) at room temperature are shown in **Fig. 2a**. The modulus of elasticity was found to be around 0.9 ± 0.2 GPa. Isomalt behaved mostly like a solid at < 50 °C. Upon mild heating to just above about 50°C, the isomalt substrates softened and could be mechanically deformed by manually applied forces. Furthermore, this deformation could be retained (mechanical memory) by cooling the substrate after deformation. Isomalt substrates were viscoelastic in the temperature range of 50 °C to 130 °C. Beyond 130 °C and upto 200 °C, isomalt behaved like a fluid. The viscosity of isomalt was studied using a falling ball viscometer and was found to vary from about 50 Ns/m² at 100 °C to 25 Ns/m² at 150 °C (**Fig. 2b**). The boiling point of isomalt was observed to be about 220 °C.

*2.1.3. Optical Properties of Isomalt.* UV-vis spectroscopy of isomalt substrates, showed 70% transmittance for a 500 μm thick sample at a wavelength of 550 nm (**Fig. 2c**) with wavelengths < 300 nm being absorbed (**Fig. 2d**). Time resolved fluorescence spectroscopy showed fluorescence at an excitation wavelength of 270 nm (**Fig. 2e**).

## 2.2 Development of Inductors

Inductors on Isomalt-Sugarin substrates were constructed using three techniques – first as an in-plane 2-dimensional (2-D) coil on the isomalt-sugarin substrate, second by metal deposition on a sculpted 3-D isomalt spiral and third as a spiral metal deposition on an iron embedded isomalt core.

*2.2.1. Planar 2D Inductors.* The first architecture for inductors was planar (2-D) Ag square coil with 6 turn, 1000 μm width and gap, 1 inch outer diameter (**Fig. 3a**). Impedance spectrum showed significant parasitic resistance of about 40 Ω with the inductive component of approximately 300 nH dominating at >5 MHz (**Fig. 3a**). The Modified Wheeler approximation predicts an inductance of 490 nH [37]. The advantages of this architecture are the simplicity of the build and the possibility of transfer to soft substrates (**Fig. 3a**). The disadvantage of the architecture is the relatively low inductance and high parasitic resistance.

*2.2.2. Helical 3D Inductors.* Taking advantage of the mechanics of isomalt, 3-D sculpted coil inductors were constructed by drawing isomalt from a melt using capillary forces. The melt was spooled into a helical wire onto a glass rod that was simultaneously rotated and translated. After cooling to room temperature the spool was coated with sugarin followed by Ag metallization (**Fig. 3b**). **Fig. 3b** shows the impedance spectrum of the inductors fabricated (7 turns, 1.6 cm long, 4 mm inner radius) with a ferrite core. Here too the parasitic resistance dominates at low frequencies with the inductance being 0.3 μH to 3 μH. The theoretical inductance for a solid wire coil of similar dimensions with a ferrite core would be 6 µH. The advantage of the 3-D sculpted coil is that it permits the use of a high permeability core. On the downside, the inductor is far harder to build and uniform metallization is more challenging. Moreover, the parasitic resistance is still significant.

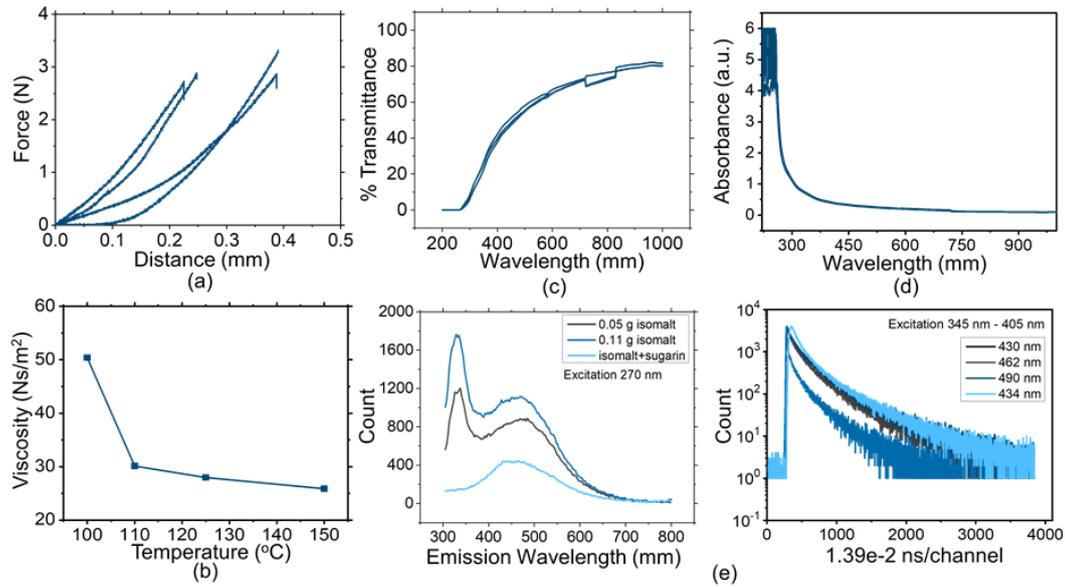

**Figure 2.** Mechanical and optical properties of isomalt-sugarin substrates.(a) Flexural strength measurements of isomalt substrate under UTM (b) Viscosity of isomalt as a function of temperature using falling ball viscometer (c) Transmittance of isomalt substrates via UV-vis spectroscopy (d) Absorbance of isomalt substrates via UV-vis spectroscopy (e) Photoluminesence studies of isomalt.

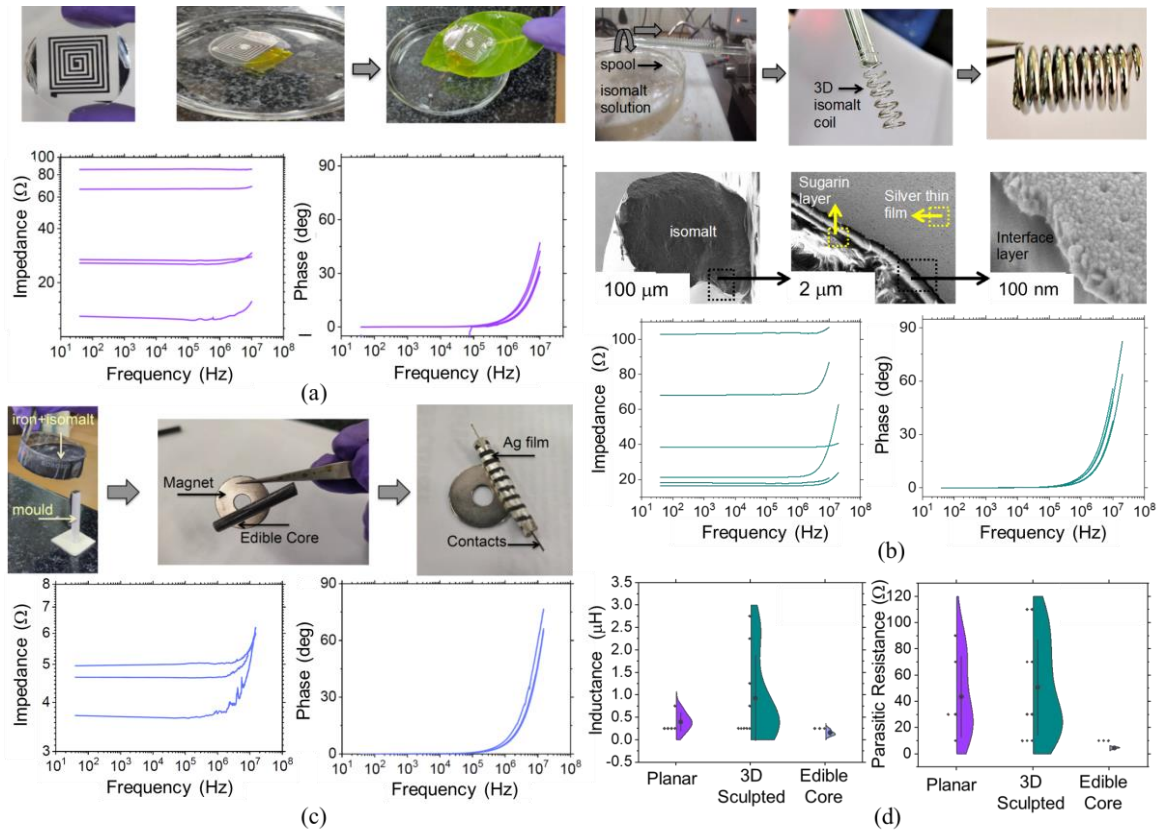

**Figure 3.** (a) 2D planar coil inductor and impedance spectrum, (b) helical coil inductors with an air core, its cross-section and impedance spectrum with a ferrite core, (c) inductors on a high permeability edible core and impedance spectrum, (d) comparison of the three architectures.

*2.2.2. 3D Inductors with a High Permeability Edible Core.* The ferrite core used in **Fig. 3b** is not edible. To construct the edible core, 100 mg of 5 µm Fe particles were mixed into 20 g of isomalt melt. The iron doped isomalt was shaped into a cylindrical core of 3 cm length and 3 mm diameter (weight of core 1.1 g, weight of iron 5mg). Ag interconnects of 2mm width and 1 mm pitch were conformally deposited directly on the core to avoid air gaps (**Fig. 3c**). The impedance spectrum of the edible core inductors is shown in **Fig. 3c**. The wider track width reduces the parasitic resistance, and the inductance is about 300 nH. The relative permeability of the core was calculated to be about 15. **Fig. 3d** compares the measured inductance and parasitic resistance across the three topologies.

## 2.3 Development of Capacitors

Parallel plate capacitors were designed with 100 nm Ag as top and bottom electrode and spin-coated sugarin as dielectric (relative dielectric constant 1.6). Typical dielectric thickness (± 20\% error) as a function of spin coating speed is shown in **Fig. 4**. The impedance spectrum of 1.0 cm by 1.0 cm capacitors with the dielectric deposited at 3000 r.p.m is also shown. The parasitic parallel resistance becomes important at higher frequencies.

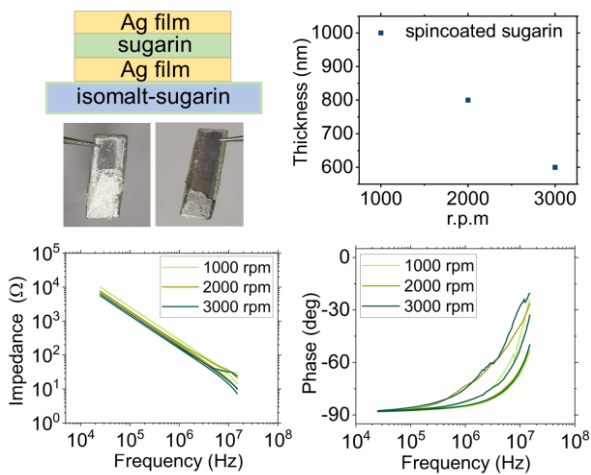

**Figure 4.** Parallel plate capacitors

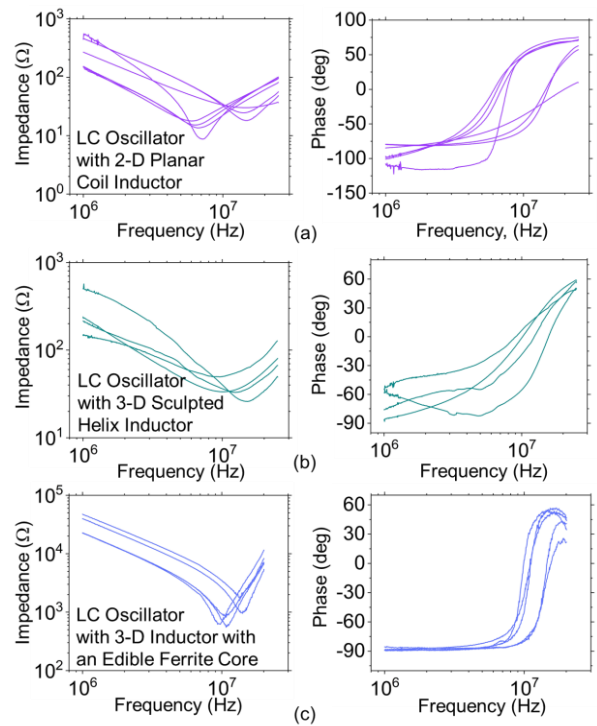

**Figure 5.** Impedance spectrum of series circuits using inductors based on (a) 2D planar (b) helical coil with ferrite core (c) edible core.

## 2.4 Edible Resonators

*2.4.1 Bonding Discrete Components.* The discrete inductor and capacitor were bonded by placing a sugarin drop at the junction. No metallization was done post bonding. The bond had a fracture stress of 78.4 kN/m$^2$. Via holes were made on isomalt-sugarin substrates using a hot (75 C) metal pin followed by metal sputtering.

*2.4.2 LC Circuits.* Series LC circuits were built using the discrete inductors and capacitors. The impedance spectra is as shown in **Fig. 5**.

## 3. Conclusions

LC circuits with resonant frequency close to 13.5 MHz using isomalt and sugarin were demonstrated. The fabrication of isomalt-sugarin substrates, inductors, capacitors and their integration was discussed.

**Acknowledgement**

The authors thank the Engineering to Clinical Practices Award, University of Cambridge, the Institute of Eminence Grant, Indian Institute of Science and DST BTDT Grant for funding. The authors thank the Centre for Nanoscience and Engineering, Indian Institute of Science for us of the characterization facilities. The authors thanks Prof. Rebecca Fitzgerald of the Medical Research Centre, University of Cambridge for valuable discussions. The authors thank Prof. Sai Siva Gorthi and Mr. Rajesh. S for fluorescence microscope imaging, Mr. Virenda Parab for SEM images, Dr. Santhosh.Bhargav and Prof. Ananth Suresh for micro newton force sensor for viscoelastic measurements. Sanjiv Sambandan thanks the DBT Cambridge Lectureship for permitting a joint appointment between the Indian Institute of Science and the University of Cambridge.